\documentstyle[12pt,fleqn]{article}
\begin{document}

\begin{titlepage}
\vspace*{-62pt}
\begin{flushright}
FERMILAB--PUB--93/098--A\\
April 1993
\end{flushright}
\vspace{1.5cm}
\begin{center}
{\Large \bf Hydrodynamic Detonation Instability in\\
Electroweak and QCD\\Phase Transitions}\\
\vspace{.6cm}
\normalsize
{Mark Abney$^*$
\vspace{12pt}

{\it

Department of Astronomy and Astrophysics, The Enrico Fermi
Institute,\\
The University of Chicago, Chicago, Illinois~~~60637\\ and\\

NASA/Fermilab Astrophysics Center,\\
Fermi National Accelerator Laboratory, Batavia, Illinois~~~60510}
\vspace{18pt}
}

\end{center}

\vspace*{12pt}

\baselineskip=24pt

\begin{quote}
\hspace*{2em}

The hydrodynamic stability of deflagration and detonation
bubbles for a first order electroweak and QCD
phase transition has been discussed recently
with the suggestion that detonations are stable.
We examine here the case of a detonation more carefully. We find that
in front of the bubble wall perturbations do not grow with time, but
behind the wall modes exist which grow exponentially.
We briefly discuss the possible meaning of this 
instability.

\vspace*{12pt}

PACS number(s): 98.80.Cq, 95.30.Lz

\vspace*{12pt}

\noindent
\small $^*$email: abney@rainbow.uchicago.edu

\end{quote}

\end{titlepage}

\baselineskip=24pt

A first order phase transition
involves the nucleation of a bubble of one phase within a medium
of the original phase. If the bubbles are large enough, they expand,
collide and coalesce until the original phase has been completely
replaced by the new phase. Though many of the details of this process
are not definitively known---the expansion rate of the bubbles for
instance---understanding the dynamics of such transitions may
provide valuable insights into the conditions of the early universe.

Of particular interest are the electroweak (EW) and QCD phase transitions
and the possible ramifications towards the generation of a baryon
asymmetry in the EW case and the concentration of baryons in the QCD case.
If the EW phase transition is first order, the possibility of
baryon asymmetry generation arises \cite{kuzm}. Furthermore, if
the asymmetry is
created by interaction with the bubble wall,
it becomes important to understand
in detail the shape and structure of the wall \cite{dine}.
A first order QCD phase transition, on the other hand,
may result in baryon concentrating
effects \cite{witt}.
Crucial to this is the concept of phase separation \cite{link}.
The generation of a baryon concentration
may depend on whether, and how effectively, the two phases
mix during the period of bubble expansion.
The stability of the bubble
wall and the possible existence of
turbulence are thus important factors
when considering the effects of these phase transitions.

Recent investigations \cite{link,kam,huet} of the
hydrodynamic stability of the bubble wall
restrict their attention to the case
of a deflagration front, namely a bubble wall
which is propagating at a speed slower than sound
relative to the old phase.
Recent estimates support the assumption that
the bubble wall will propagate subsonically \cite{dine1}.
Even if these estimates are correct, however,
there may exist situations which allow
for the supersonic propagation of the
bubble, {\it i.e.,\ }a detonation front.
It was suggested by Kamionkowski and Freese \cite{kam}
that the existence of an instability in the deflagration
front could result in the acceleration
of the front until it becomes a detonation,
a phenomenon which is observed
in laboratories studying combustion \cite{zeld}.
The paper by Huet {\it et al.\ }\cite{huet} analyzed
the dispersion relation obtained for
a perfect relativistic fluid
and concluded that there
could be no perturbations which grow
in time in the detonation case.
They, however, only examine the region in front of the
expanding bubble and not the region behind it.
In this paper we employ the standard linear stability
analysis used by others \cite{link,huet,dyak,land}
to examine the possible existence of instabilities
both in front of and behind the bubble wall
in the detonation case.

Detonations and deflagrations consist of two different
phases separated by a transition
region referred to as the bubble wall. A detonation front
propogates at a velocity greater
than the local speed of sound relative to the original phase.
Behind the wall, the medium has
a velocity equal to or less than the speed of sound in the
second phase (see Steinhardt's article \cite{stein}
for a discussion of relativistic detonation and shock waves).
We choose to move into
the frame of the moving wall and position it at $x=0$.
To the left of the wall ($x<0$) is the original
({\it e.g.,\ }quark) phase, while to the right
($x>0$) is the new ({\it e.g.,\ }hadron) phase.
In region 1 ($x<0$) the fluid has a positive velocity greater
than that of sound ($v_1>c_{s1}>0$). In region 2 ($x>0$) the fluid
has a positive velocity less than that of sound
($0<v_2<c_{s2}$)\cite{stein}.
The situation is illustrated in Figure 1.
Because of the different velocities, the behavior
of perturbations and their effect on the bubble wall will differ.
Since perturbations
cannot travel faster than the speed of sound, we expect that
in region one they should be ``swept away'' as the
fluid passes through the wall. This is in agreement with
references \cite{kam,huet}.
The situation behind the wall is considerably different, there being
no {\it a priori\/} reason why one should expect perturbations
to decay with time.

We first consider the behavior of a relativistic perfect fluid
in the two separate regions. Using the metric $g=(+,-,-,-)$ the
stress-energy of a perfect fluid is
$T^{\mu\nu}=wu^{\mu}u^{\nu}-pg^{\mu\nu}$
where we have taken $c=1$, with $e$ the energy density,
$p$ the pressure,
and $w=e+p$ the enthalpy density.
The equations of motion are:
$$
  u^\mu\partial_\mu p + c_s^2w\partial_\mu u^\mu=0
$$
$$
  wu^\mu\partial_\mu u_\nu - \partial_\nu p+u_\nu u^\mu\partial_\mu p=0
$$
where $u_\mu=(\gamma,\gamma\vec v)$ and $\gamma=1/\sqrt{1-v^2}$.
We take perturbations in the
velocity and pressure of the fluid about a constant solution:
$$
\begin{array}{l}
	p=p_0+\delta p \\
	\vec{v}=\vec v_0+\delta\vec v \\
	\vec v_0=v_0\hat{x}\\
	\delta\vec v=\delta v_x\hat{x}+\delta v_y\hat{y}.
\end{array}
$$
Keeping terms up to first order in $\delta p$ and $\delta\vec v$,
we can write the equations of motion as:
\begin{eqnarray}
	(1-c_s^2v_0^2)\frac{\partial}{\partial t}\delta p
	+v_0(1-c_s^2)\frac{\partial}{\partial x}\delta p
	+wc_s^2(\frac{\partial}{\partial x}\delta v_x
	+\frac{\partial}{\partial y}\delta v_y)=0	\\
	\frac{v_0}{w\gamma^2}\frac{\partial}{\partial t}\delta p
	+\frac{\partial}{\partial t}\delta v_x
	+\frac{1}{w\gamma^2}\frac{\partial}{\partial x}\delta p
	+v_0\frac{\partial}{\partial x}\delta v_x=0	\\
	\frac{\partial}{\partial t}\delta v_y
	+v_0\frac{\partial}{\partial x}\delta v_y
	+\frac{1}{w\gamma^2}\frac{\partial}{\partial y}\delta p=0
\end{eqnarray}
where (1) comes from the conservation of energy and (2)--(3) are
the relativistic Euler's equations.
It is more convenient to write this system as the matrix equation:
\begin{equation}
  {\bf A_t}\frac{\partial}{\partial t}\vec W
  +{\bf A_x}\frac{\partial}{\partial x}\vec W
  +{\bf A_y}\frac{\partial}{\partial y}\vec W =0  \label{matrix}
\end{equation}
where
$$
  \vec W=\left(  \begin{array}{c}
		   	\delta p	\\
			\delta v_x	\\
			\delta v_y
		\end{array}  \right)
$$
$$
  {\bf A_t}=\left(  \begin{array}{ccc}
			1-c_s^2v_0^2	& 0	& 0	\\
{\displaystyle\frac{v_0}{w\gamma^2}}	& 1	& 0	\\
			0		& 0	& 1
		    \end{array}  \right) \mbox{\ \ }
  {\bf A_x}=\left(  \begin{array}{ccc}
			v_0(1-c_s^2)	& w c_s^2 &0	\\
{\displaystyle\frac{1}{w\gamma^2}}	& v_0	& 0	\\
			0		& 0	& v_0
		    \end{array}  \right) \mbox{\ \ }
  {\bf A_y}=\left(  \begin{array}{ccc}
			0		& 0	& w c_s^2 \\
			0		& 0	& 0	\\
{\displaystyle	\frac{1}{w\gamma^2}}	& 0	& 0
		    \end{array}  \right).
$$
This system may be solved to obtain a solution of the form:
$$
  \vec W(x,y,t)=\vec F(x) e^{-i(\omega t+ky)}
$$
where
$$
  \vec F(x)=\sum_{j}a_j e^{-iq_jx}\vec R_j.
$$
The $a_j$ are constants, and $\vec R_j$ are eigenvectors. The $q_j$
are found by solving the characteristic equation for (\ref{matrix}).
Doing this, we obtain the dispersion relation
\begin{equation}
  (qv_0+\omega)\left[
		\frac{1}{c_s^2}(qv_0+\omega)^2-(q+v_0\omega)^2
		-(1-v_0^2)k^2
	     \right]=0 \label{dspeqn}
\end{equation}
The solutions for q are:
\begin{equation}
   q_1 =\frac{-\omega}{v_0}
\end{equation}
\begin{equation}
   q_{2,3} = \frac{1}{v_0^2-c_s^2}\left[(c_s^2-1)v_0\omega\pm
      c_s(1-v_0^2)\sqrt{\omega^2+\frac{v_0^2-c_s^2}{1-v_0^2}k^2}\,\right].
\label{disp}
\end{equation}
These, (\ref{dspeqn})--(\ref{disp}),
are the same as the equations obtained in reference \cite{huet}
equations (32), (34)--(36).
It will be of future interest to note that in the case
where $v_0=c_s$ we obtain two solutions, $q_1$ from above and
\begin{equation}
  q_2=\frac{-(1+c_s^2)}{2c_s}\omega+\frac{c}{2}\frac{k^2}{\omega}.
  \label{cjq}
\end{equation}
Thus, the solution for perturbations in a perfect relativistic
fluid is:
\begin{equation}
  \vec W(x,y,t)=\left(a_1\vec R_1e^{-iq_1x} +
	a_2\vec R_2e^{-iq_2x} +a_3\vec R_3
	e^{-iq_3x}\right)e^{-i(\omega t+ky)}  \label{soln}
\end{equation}
where
$$
  \vec R_1= \left(	\begin{array}{c}
				0 \\ 1 \\{\displaystyle\frac{q_1}{k}}
			\end{array}
	    \right)\mbox{ and }
  \vec R_j= \left(	\begin{array}{c}
			  1 \\
			{\displaystyle\frac{-(q_j+v_0\omega)}
			{w\gamma^2(\omega+v_0q_j)}}\\
			{\displaystyle\frac{-k}{w\gamma^2(\omega+v_0q_j)}}
			\end{array}
	    \right)
  \mbox{  for }  j=2,3.
$$

We now wish to examine whether there exist any unstable modes
of $\vec W$ ({\it i.e.,} modes where Im~$\omega>0$) which
obey the boundary conditions $\vec W\rightarrow 0 \mbox{ as }
x\rightarrow \pm \infty.$
First, consider region one ($x<0$). In order to satisfy the
boundary condition as $x\rightarrow -\infty$, we must require either
Im~$q_j>0$ or $a_j=0$. However, with some
algebra, we can show from (\ref{disp})
that if Im~$\omega >0$ then Im~$q_j<0$ and therefore $a_j=0$ for
$j=1,2,3$.
That is, perturbations which grow in time cannot exist in
front of the bubble wall. This is the conclusion reached in
reference \cite{huet}.
Let us next consider what occurs in region two
($x>0$). Here, the $x\rightarrow +\infty$ boundary condition
requires either Im~$q_j<0$ or $a_j=0$. We find, if Im~$\omega>0$ then
Im~$q_{1,2} <0$ and Im~$q_3>0$.
Thus, behind the wall we need require only that $a_3=0$.
That is, since we can satisfy the $x\rightarrow +\infty$
boundary condition without requiring $a_{1,2}=0$, the solution for
the perturbations, equation (\ref{soln}), is not identically zero,
{\it i.e.,\ }perturbations behind the bubble wall
that grow in time may exist.

The above work, however, is not enough to prove that
instabilities do, in fact, occur. It is still necessary to
impose the constraints of the boundary conditions across
the bubble wall to determine whether the instabilities
do not contradict the relevant conservation laws.
Let us use notation where subscripts of one or two indicate
quantities in region one ($x<0$) or two ($x>0$) respectively.
Also, we assume that there exists a perturbation of the
bubble shape of the form:
$$
  \Delta (y,t)=De^{-i(\omega t+ky)}.
$$
We require
\begin{itemize}
  \item conservation of energy:
    $$ 
	w_1\gamma_1^2v_1=w_2\gamma_2^2v_2  
    $$ 
  \item conservation of momentum:
    $$ 
	w_1\gamma_1^2v_1^2+p_1=w_2\gamma_2^2v_2^2+p_2-
	\sigma \left(\frac{\partial^2}{\partial y^2}-\frac{\partial^2}
	{\partial t^2}\right)\Delta  
    $$ 
  \item continuity of transverse velocity:
    $$ 
	v_{1y}+v_1\frac{\partial\Delta}{\partial y}=
	v_{2y}+v_2\frac{\partial\Delta}{\partial y}.  
    $$ 
\end{itemize}
Recalling that $W=0$ in region one, the above equations
may then be linearized to obtain in region two:
\begin{eqnarray}
  \delta p &=& \frac{1}{\Gamma_-}\left[-2\gamma_2^2w_2v_2\frac{v_1-v_2}{v_1}
		(-i\omega)+\sigma(\omega^2-k^2)\right]\Delta \label{energy}\\
  \delta v_x &=&\frac{\Gamma_+}{\Gamma_-}\left(\frac{v_1-v_2}{v_1}\right)
		(-i\omega)\Delta+\left(1-\frac{\Gamma_+}{\Gamma_-}\right)
		\frac{\sigma(\omega^2-k^2)}{2\gamma_2^2w_2v_2}\Delta  \\
  \delta v_y &=& (v_1-v_2)(-ik)\Delta  \label{tvel}
\end{eqnarray}
where $\Gamma_\pm=1\pm\gamma_2^2\theta_2v_2^2$ and
$\theta_2=1+1/c_{s2}.$
Then (\ref{energy})--(\ref{tvel}) give us
the perturbations just interior to the bubble,
{\it i.e.,\ }as $x\rightarrow 0^+$. Since we
require $a_3$ to be zero for $x>0$ we are left with $a_1$, $a_2$,
and $D$ as undetermined constants. By matching the general solution
of perturbations for $x>0$, equation (\ref{soln}), with
equations (\ref{energy})--(\ref{tvel}), we obtain three
equations for these three unknowns. We will always have a solution
for such a system provided that the determinant of the
coefficient matrix vanishes. Let us write equations
(\ref{energy})--(\ref{tvel}) as
\begin{equation}
  \vec Y \Delta
  \equiv\left(\begin{array}{c}
		\delta p\\ \delta v_x\\ \delta v_y
	\end{array}\right).  \label{perts}
\end{equation}
Matching solutions (\ref{soln}) with (\ref{perts}) we get:
$$
  \vec W(0^+,y,t)=\left(a_1\vec R_1+a_2\vec R_2\right)e^{-i(\omega t+ky)}
  =\vec Y De^{-i(\omega t+ky)}.
$$
This equation may be rearranged into a single $3\times 3$ matrix
equation where the columns of the matrix are the vectors $\vec R_1$,
$\vec R_2$ and $\vec Y$.
$$
  \left( \vec R_1 \left| \vec R_2 \right| \vec Y \right)
  \left( \begin{array}{c} a_1 \\ a_2 \\ D \end{array} \right)=0
$$
By taking the determinant of the above $3\times3$ matrix we
obtain an equation in $\omega$, $q_2$, $k$, $\sigma/w_2$,
$v_1$, $v_2$, and $c_{s2}$:
\begin{eqnarray}
  \lefteqn{\frac{1}{\Gamma_-v_2k(\omega+v_2q_2)}\times} \nonumber \\
    & & \left(
	  i\frac{v_1-v_2}{v_1}
		\left[
		  (\Gamma_+-2v_2^2)\omega^3+v_2(\Gamma_+-2)q_2\omega^2+
		  (2v_2-\Gamma_-v_1)v_2k^2\omega-\Gamma_-v_1v_2^2q_2k^2
		\right]
	\right.
  \nonumber \\
    & & \left.
	  +\frac{\sigma(\omega^2-k^2)}{2\gamma_2^2w_2v_2}
		\left[
		  (\Gamma_+ -\Gamma_-(1+2v_2^2))\omega^2+
		  (\Gamma_+-3\Gamma_-)v_2q_2\omega
		  +2\Gamma_-v_2^2k^2
		\right]
	\right)=0  \label{geneq}
\end{eqnarray}
Combining equations (\ref{disp}) and (\ref{geneq}) and solving
for $\omega$ we can eliminate $q_2$ and obtain
an equation for $\omega$ as a function of $v_1$, $v_2$, $c_{s2}$,
$\sigma/w_2$ and $k$. By studying this equation we can determine
if there exist any cases where $\omega$ has a positive imaginary
solution, thereby establishing that the detonation has a
self-sustaining instability.

Of particular relevance is the case of a Chapman-Jouget detonation.
This occurs when the velocity behind the wall is equal to the speed
of sound ($v_2=c_{s2}$). Steinhardt showed that any detonation with
spherical symmetry must be of the
Chapman-Jouget type \cite{stein}. It has also
been postulated that any ``naturally'' occuring detonation will
meet this condition~\cite{land}.
In order to
examine this case we use the value for $q_2$ given in equation (\ref{cjq}).
It is interesting to look at the case where the surface tension
is zero ($\sigma=0$); there are four solutions
to $\omega$, two of which are positive imaginary. The
solutions are:
\begin{eqnarray*}
  \omega=\pm i \sqrt{\frac{c_{s2}v_1}{2-c_{s2}^2}}\,c_{s2}k  \\
  \omega=\pm i\sqrt{\frac{1}{1-c_{s2}}}\,c_{s2}k
\end{eqnarray*}

The full equation with surface tension cannot be solved analytically.
Instead we look at the behavior of $\omega$ under different limits.
In the long wavelength limit ($k\rightarrow 0$), keeping terms
up to second order in $k$, equation (\ref{geneq}) becomes:
\begin{eqnarray}
\lefteqn{ \frac{3\sigma(1-c_s^2)}{2w_2c_s}(1-c_s^4)\omega^5
  +i\frac{v_1-c_s}{v_1}(c_s^4-3c_s^2+2)\omega^4
+\frac{\sigma(1-c_s^2)}{w_2c_s}(-c_s^4+\frac{3}{2}
  (c_s^2-1))k^2\omega^3} \nonumber \\
 && \mbox{}+i(-c_s^3v_1-c_s^2+c_sv_1+2)c_s^2\frac{v_1-c_s}{v_1}k^2\omega^2=0
\end{eqnarray}
Notice that the quantity $\sigma/w_2$ provides a natural length scale.
In Figure 2, the behavior of Im~$\omega$ was
plotted as a function of $k$ with
values of $v_1=1.5\,c_s$ and $c_s=1/\sqrt{3}$.
The particular choice for $v_1$ is arbitrary;
the behaviour of Im~$\omega$ is not altered
significantly for different values.
We have chosen a solution which matches
with an unstable mode in the zero surface tension case,
and we see that
the rate of expansion increases slightly slower than linearly with
the wave number, an effect of the surface tension.
The discontinuity which appears as $k$ increases is an artifact
of the small $k$ approximation; its location
approaches the origin as the surface tension is increased. That is,
with a larger surface tension we must go to a larger wavelength
in order to insure that the $k\rightarrow 0$ approximation is valid,
as we expect.

Next, consider the short wavelength limit. As $k$ goes to infinity
only higher orders of $k$ contribute and we get:
\begin{eqnarray}
  \frac{\sigma(1-c_s^2)}{w_2c_s}(-c_s^4+\frac{3}{2}(c_s^2-1))k^2\omega^3
  +i(-c_s^3v_1-c_s^2+c_sv_1+2)c_s^2\frac{v_1-c_s}{v_1}k^2\omega^2\nonumber\\
  \mbox{}-\frac{\sigma(1-c_s^2)^2}{2w_2}c_sk^4\omega
  +i(v_1-c_s)c_s^5k^4=0.
\end{eqnarray}
Figure 3 shows Im~$\omega$ as a function of $k$ with the same
values as in Figure 2. Here we notice that as $k$ gets larger
Im~$\omega$ quicky approaches a constant value of approximately
$0.144\,w_2/\sigma$. Unlike in the deflagration case, where the surface
tension results in there being a lower limt cutoff in the wavelength
of instabilites, with a Chapman-Jouget detonation the growth rate
does not drop to zero with shorter wavelengths, but rather reaches
a maximum positive value.

In order for these instabilities to be dynamically
relevant the mode with the fastest growth time scale
must be less than the time scale associated with the phase
transition. The QCD transition lasts a time
$t_H\sim10^{-5}\,{\rm s}$ and has a value of
$\sigma/w_2\sim 1\,{\rm fm}$ \cite{link}. We showed above that
Im~$\omega$ reached a maximum value of $\sim 0.14\,w_2/\sigma$,
corresponding to a time of $\sim 2.3\times 10^{-23}\,{\rm s}$.
Thus, there is ample time for the instabilities to mature.

The scales associated with the EW phase transition are much
smaller. In this case, using the formulas and parameter
values as given in reference \cite{kaj},
we have $\sigma\sim 0.09 T_c^3$ and
$w_2\sim 40 T_c^4$ where $T_c$ is the critical temperature.
With a critical
temperature of $150\,{\rm GeV}$ we get
$\sigma/w_2\sim 3\times 10^{-6}\,{\rm fm}$. This leads to
a maximum value of Im~$\omega$ of
$\sim 5\times 10^4\,{\rm fm}^{-1}$ corresponding
to a time of $\sim 6.7\times 10^{-29}\,{\rm s}$. Since the
phase transition lasts a time $\sim 0.005 t_H$ \cite{kaj},
where the Hubble time is $t_H\sim 10^{-11}\,{\rm s}$,
we see that there is sufficient time for the instabilities
to grow.

Our general picture of the instabilities, then, is the following.
For very large wavelengths the instabilities grow only very
slowly with time, vanishing as $\lambda$ approaches infinity.
As the wavelength decreases
Im~$\omega$ rapidly increases with
smaller $\lambda$ until a maximum rate is reached.
Thus, unstable modes exist at all wavelengths.
Even though these calculations were done
for a Chapman-Jouget type of detonation, it is unlikely
that the existence and time scale of the instabilities
is so sensitive to the velocity behind the wall that
this picture would be very much altered should $v_2<c_s$.
Additional work beyond the scope of this report would
need to be done, however, to verify this assertion.

Note that these instabilities do not exist in front of the
bubble wall, but rather behind it and at its surface.
What, then, are the ramifications on the bubble wall and
the fluid inside it?
To answer this, let us examine similar types of instabilites which
exist in the laboratory. D'yakov and others \cite{dyak} have
calculated the stability condition for shock waves in a
classical non-relativistic fluid. Though a shock wave and
detonation are not identical, the fluid dynamics are quite
similar. It is possible to establish conditions in the
laboratory which violate this stability requirement; that is,
under certain circumstances there are unstable
modes which exist on the surface of and behind the shock wave.
Thompson {\it et al.\ }\cite{thomp} have recently
done experiments which violate this
shock stability condition. In their report
they show a series of photographs
where we see shock waves  transform from planar to
a billowy cloud-like surface as the stability condition is violated.
The surface of the wave is travelling
faster than the speed of sound, and the instabilities
do not propogate forward from the shock.
The transformation of the shock surface from
planar to irregular is described by them as a ``transition to turbulence''
of the fluid behind the shock.

It is possible that a similar situation
exists with the instabilites described above. The growth of perturbations
behind the bubble wall may result in turbulence and a highly
irregular surface. Such effects may entail an alteration in
the picture of the phase transition proceeding through the
growth of uniform spherical bubbles.
This may very well be relevant when considering the possibility
of baryon generation and concentration in the EW and QCD
phase transitions respectively.

Whether detonations actually do arise, though, is
not yet known. A more highly first order transition would
likely result in an increased possibility of detonations.
Another mechanism could be the existence of instabilities
in deflagrations as desribed in reference \cite{kam}.
The above analysis, however, is limited to the linear
regime and does not take into account any non-linear effects that
may arise. The result of such effects may be to stablize the
perturbations, as has been observed with deflagrations in the laboratory
\cite{zeld}. It remains
to be seen whether non-linearities are indeed important.

I would like to thank E. Kolb and G. Starkman for helpful discussions
and comments. This work was supported in part by the University of
Chicago Physics Department, the DOE and NASA grant NAGW-2381 at
Fermilab.

\pagebreak

\Large
{\bf Figure captions}
\normalsize
\bigskip
\baselineskip=24pt

{\bf Figure 1}: Schematic of the bubble wall in the frame of the
wall. The wall is located at $x=0$ with the old phase
({\it e.g.,\ }quarks) in the half-plane $x<0$ and the
new phase ({\it e.g.,\ }hadrons) in the half plane $x>0$.

\medskip

{\bf Figure 2}: Perturbation growth rate (Im~$\omega$) as a function
of wave number ($k$) in the limit $k\rightarrow 0$, plotted with
parameter values $c_s=1/\sqrt{3}$, $v_1=1.5\,c_s$
in units of $w_2/\sigma$.

\medskip

{\bf Figure 3}: The same as in Figure 2, but in the limit
$k\rightarrow\infty$.

\end{document}